\documentclass[twocolumn,prl,tightenlines,superscriptaddress,floatfix]{revtex4-2}

\usepackage{hyperref}
\usepackage{amsmath,amssymb}
\usepackage[utf8]{inputenc} 			% Encodage UTF-8
\usepackage[T1]{fontenc}				% Encodage du fichier de sortie
\usepackage{graphicx}
\usepackage{float}
\usepackage[usenames,dvipsnames]{xcolor}
\usepackage{tikz}
\usepackage{ulem}

\newcommand{\grad}{\nabla}

\renewcommand{\vec}[1]{\mathbf{#1}}
\newcommand{\Gvec}[1]{\boldsymbol{#1}}

\newcommand*\colvec[3][]{
    \begin{pmatrix}\ifx\relax#1\relax\else#1\\\fi#2\\#3\end{pmatrix}
}

\begin{document}

\title{Dynamic clustering of passive colloids in dense suspensions of motile bacteria}

\author{Shreyas Gokhale$^{\ddag,}$}
\affiliation{Department of Physics, Massachusetts Institute of Technology, Cambridge, MA 02139, USA}

\author{Junang Li$^{\ddag,}$}
\affiliation{Department of Physics, Massachusetts Institute of Technology, Cambridge, MA 02139, USA}

\author{Alexandre Solon}
\affiliation{Sorbonne Universit\'e, CNRS, Laboratoire de Physique Th\'eorique de la Mati\`ere Condens\'ee, LPTMC, F-75005 Paris, France}

\author{Jeff Gore}
\email[Corresponding author: ]{gore@mit.edu}
\affiliation{Department of Physics, Massachusetts Institute of Technology, Cambridge, MA 02139, USA}

\author{Nikta Fakhri}
\email[Corresponding author: ]{fakhri@mit.edu}
\affiliation{Department of Physics, Massachusetts Institute of Technology, Cambridge, MA 02139, USA}

\date{\today}
\begin{abstract}
Mixtures of active and passive particles are predicted to exhibit a variety of nonequilibrium phases. Here we report a dynamic clustering phase in mixtures of colloids and motile bacteria. We show that colloidal clustering results from a balance between bond breaking due to persistent active motion and bond stabilization due to torques that align active particle velocity tangentially to the passive particle surface. Furthermore, dynamic clustering spans a broad regime between diffusivity-based and motility-induced phase separation that subsumes typical bacterial motility parameters. 
\end{abstract}
\maketitle

Collective self-organization of self-propelled, or active, particles is a vibrant topic of research in statistical physics~\cite{marchetti2013hydrodynamics,bowick2021symmetry}. Active matter is known to exhibit a diverse array of nonequilibrium phenomena such as flocks~\cite{bricard2013emergence}, living crystals~\cite{palacci2013living,tan2021development}, active nematics~\cite{narayan2007long,sanchez2012spontaneous,duclos2020topological,copenhagen2021topological}, turbulent phases ~\cite{wensink2012meso}, whorls~\cite{bililign2021chiral} and nonreciprocal interactions ~\cite{zhang2021active}. Over the last decade, an emerging area in active matter physics has been the interactions between active and passive particles~\cite{bechinger2016active}. Mixtures of active and passive particles have been studied in the context of enhanced colloidal diffusion in active environments~\cite{wu2000particle,patteson2016particle,peng2016diffusion}, transport of colloidal cargo by bacteria~\cite{koumakis2013targeted,vaccari2018cargo}, pairwise interactions between colloids in bacterial baths~\cite{angelani2011effective,liu2020constraint}, and the effect of active dopants on colloidal fluids~\cite{kummel2015formation} and crystals~\cite{ramananarivo2019activity}. Further, theoretical and numerical work on binary mixtures of active and passive particles predicts novel forms of nonequilibrium self-assembly such as segregation based on differences in motility~\cite{mccandlish2012spontaneous,stenhammar2015activity,dolai2018phase} or diffusivity~\cite{grosberg2015nonequilibrium,weber2016binary}. However, a fundamental understanding of the mechanisms through which the nonequilibrium drive imposed by active particles guides the self-assembly of active-passive mixtures is lacking. Here, we address this challenge by investigating the dynamics of passive particles in an active bath using experiments on mixtures of colloids and motile bacteria, as well as Brownian dynamics simulations.

\begin{figure}
  \centering
  \includegraphics[width=1\linewidth]{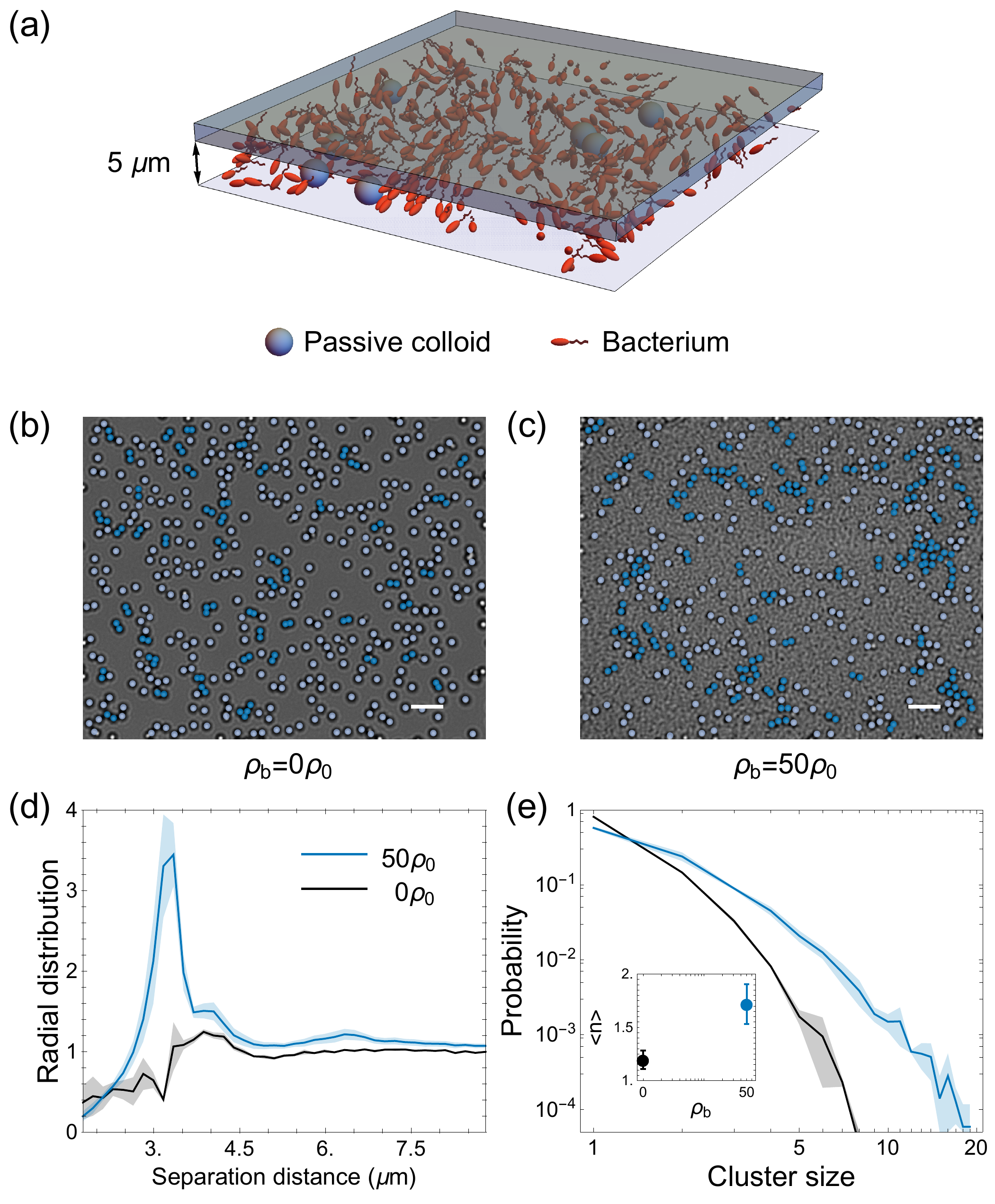}
  \caption{{\bf Passive colloids undergo dynamic clustering in suspensions of motile bacteria.} (a) Schematic of the experimental setup. Top and bottom coverslips are separated by spacer beads to ensure that the chamber thickness is $5\,\mu$m. The bottom coverslip is permeable to oxygen which allows bacterial motility to remain unchanged well beyond the duration of the experiment. (b-c) Representative snapshots of colloids in the absence of bacteria (b) and at a high bacterial density $\rho_{\rm b}=50\rho_0$ (c). Clusters with more than one particle are shown in dark blue. The scale bar is $10 \mu$m. (d) Radial distribution function $g(r)$ in the absence of bacterial (black) and high bacterial density (blue). (e) Cluster size distribution in the absence of bacterial (black) and high bacterial density $\rho_{\rm b}=50\rho_0$ (blue). Inset shows the mean cluster size computed from data shown in the main plot. Shaded regions in (d) and (e) and error bars in inset to (e) are standard errors of the mean across three independent experiments.}
  \label{fig:1}
\end{figure}

Our experiments employ a quasi-2D geometry, with passive silica colloids of diameter $\sigma=3.2\,\mu$m confined between two coverslips separated by a gap of $5\,\mu$m using spacer particles (Fig.~\ref{fig:1}a). We held the area fraction of passive colloids constant at $\phi_{\rm p}=0.15$, and systematically varied the bacterial density $\rho_{\rm b}$. We used motile bacteria of the species \textit{Pseudomonas aurantiaca} as our choice of active particles and restricted our experiments to bacterial densities below the onset of active turbulence~\cite{wensink2012meso} (See supplemental information for details about the strain~\cite{SI}). The bacteria were grown for 16 hours at $30^{\circ} C$ in yeast extract and soytone broth, whereupon they reached a saturation density $\rho_0=3\times 10^9$ colony forming units (CFU) / mL. Throughout the manuscript, we report the bacterial density $\rho_{\rm b}$ in units of $\rho_0$. 

Our experiments show that while colloids are distributed homogeneously in the absence of bacteria, they undergo significant clustering at high bacterial densities (Fig.~\ref{fig:1}b-c, Video~S1). We first verified that the observed clustering is not induced by metabolites produced by the bacteria (Fig.~S1)~\cite{SI}.  Clustering results in a strong enhancement of the first peak of the radial distribution function $g(r)$ (Fig.~\ref{fig:1}d). Furthermore, the cluster size distribution $P(n)$, where $n$ is the number of colloids in a cluster, becomes significantly broader for high bacterial densities (Fig.~\ref{fig:1}e), resulting in a larger mean cluster size $\langle n\rangle=\sum nP(n)$ (Fig.~\ref{fig:1}e, inset). We observe qualitatively similar clustering with a different bacterial species {\it Escherichia coli} as well, demonstrating that the phenomenon is general and robust (Fig.~S2)~\cite{SI}.

Even though our clusters form and break on the order of seconds, the cluster size distribution does not evolve over 30 minutes, suggesting that the system is in a nonequilibrium quasi-steady state (Fig.~S3)~\cite{SI}. This is substantially different from the phase separation of active and passive particles predicted in previous theoretical and computational studies. A key parameter that determines phase behavior of active particles is the P\'{e}clet number $Pe=v_{\rm a}\tau_{\rm r}/d_{\rm a}$ where $v_a$ is the speed of active particles, $\tau_{\rm r}$ is a reorientation time scale set by rotational diffusion for active Brownian particles or tumbling time for bacteria, and $d_{\rm a}$ is the size of the active particles. At high P\'{e}clet numbers ($Pe\gg 1$), mixtures of active and passive particles are expected to undergo segregation~\cite{mccandlish2012spontaneous,stenhammar2015activity,dolai2018phase} accompanied by large inhomogeneities in active particle density, consistent with motility-induced phase separation (MIPS) in purely active systems~\cite{cates2015motility}. However, the measured  P\'{e}clet number for our bacteria ($Pe \approx 14$, Fig.~S4) is too small for MIPS-like segregation to occur~\cite{SI}. Consistent with this expectation, we do not observe any spatial inhomogeneity in bacterial density. Phase separation can also occur in mixtures of particles that differ significantly in their diffusivities~\cite{grosberg2015nonequilibrium,weber2016binary}. For active-passive mixtures, this occurs in the limit of small P\'{e}clet numbers ($Pe<1$). This limit is also not applicable to our system, as the P\'{e}clet number for our bacteria is large enough to suppress phase separation completely. The observed steady state dynamic clustering is a distinct effect that has neither been predicted nor observed before.

\begin{figure*}
  \centering
  \includegraphics[width=1\linewidth]{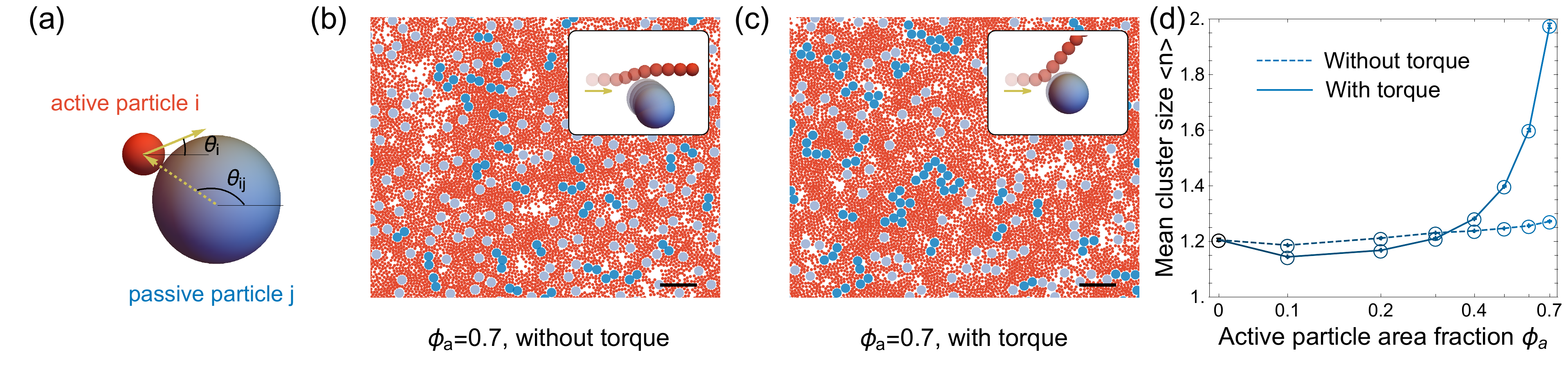}
  \caption{{\bf Simulations show that torques associated with collisions between active and passive particles stabilize steady state clustering.} (a) Schematic showing the orientation of active particles parametrized by $\theta_i$ and $\theta_{ij}$ the orientation of the relative position vector between active particle $i$ and passive particle $j$. (b-c) No clustering of passive particles is observed in the absence of aligning torques ($\Gamma=0$) (b) but substantial clustering is observed in the presence of torques ($\Gamma=25$/s) (c). Insets in (b) and (c) reveal the difference in particle trajectories during a collision, with (c) and without (b) torque. The scale bar is $10 \mu$m.(d) Mean cluster size as a function of active particle area fraction $\phi_{\rm a}$ in the absence of torque (dotted line) and presence of torque ($\Gamma=25$/s (solid line). In (b-d), the rotational diffusion coefficient  $D_{\rm r} = 1$/s. Error bars are standard errors of the mean across three independent simulations.}
  \label{fig:2}
\end{figure*}

To understand the origin of dynamic clustering, we investigated a minimal model system consisting of active and passive Brownian disks of diameter $d_{\rm a}$ and $d_{\rm p}$ respectively, in two dimensions. Both types of particles obey an overdamped Langevin equation in two spatial dimensions. The position $\vec r_i$ of passive particle $i$ follows
\begin{equation}
  \label{eq:Lpassive}
  \frac{d \vec r_i}{dt}=\mu_{\rm p} \vec F_i+\sqrt{2D_{\rm t,p}}\Gvec \xi_i
\end{equation}
where $\mu_{\rm p}$ is the mobility of a passive particle, $\Gvec \xi_i$ a 2D Gaussian white noise with unit variance and $D_{\rm t,p}$ the coefficient of translational diffusion of a passive particle. In addition, the active particles self-propel at a speed $v_0$, thus following
\begin{equation}
  \label{eq:Lactive}
  \frac{d \vec r_i}{dt}=v_0\vec u_i +\mu_{\rm a} \vec F_i+\sqrt{2D_{\rm t,a}}\Gvec \xi_i
\end{equation}
where $\mu_{\rm a}$ and $D_{\rm t,a}$ are the mobility and translational diffusion coefficient of an active particle, respectively. 
$\vec u_i=(\cos\theta_i,\sin\theta_i)$ is a unit vector giving the direction of propulsion of particle $i$. It is parametrized by an angle $\theta_i$ subject to rotational diffusion with coefficient $D_{\rm r}$ and to torques that tend to align tangentially to the surface of the passive particles
\begin{equation}
  \label{eq:Ltheta}
  \frac{d \theta_i}{dt}=-\Gamma\sum_{j}\sin(\theta_i-\theta_{ij})+\sqrt{2D_r} \eta_i
\end{equation}
where the sum runs over passive particles in contact with particle $i$ (Fig.~\ref{fig:2}a). $\Gamma$ sets the speed of alignment, $\theta_{ij}=\arg(\vec r_i-\vec r_j)$ and $\eta_i$ is a Gaussian white noise of unit variance. This type of interaction is standard to model the alignment of self-propelled rods with the surface of passive objects~\cite{chepizhko2013optimal}. Finally, the interaction $F_i$ in Eqs.~(\ref{eq:Lpassive})-(\ref{eq:Lactive}) accounts for volume exclusion. It derives from a potential so that $\vec F_i=-\grad_{r_i}U$ with
\begin{equation}
  \label{eq:pot}
  U\!=\!-\frac{k}{2}\sum_{i<j}\left(\frac{d_i+d_j}{2}-|\vec r_i-\vec r_j|\right)^2\!\!\Theta\left(\frac{d_i+d_j}{2}-|\vec r_i-\vec r_j|\right)
\end{equation}
where $d_i$ and $d_j$ assume the value $d_{\rm a}$ for active particles and $d_{\rm p}$ for passive ones. The Heaviside $\Theta$ function insures that particles interact only when they are in contact, {\it i.e.} when the distance between them is less than the sum of their radii. Eq.~(\ref{eq:pot}) imposes a soft harmonic repulsion between the particles but in practice we take a large $k=100$ that allows for little overlap. Throughout the paper, we choose $d_{\rm a} = 1\mu$m, $d_{\rm p} = 3\mu$m, $v_0 = 10\mu$m/s, $\mu_{\rm a}=(d_{\rm p}/d_{\rm a})\mu_{\rm p}=1 \text{mPa}^{-1} \cdot \text{s}^{-1} \cdot \mu \text{m}^{-1}$ ,and $D_{\rm t,a} =(d_{\rm p}/d_{\rm a})D_{\rm t,p} = 0.15 \mu$m$^2$/s.

Simulations of our model system with $\Gamma=0$ show that motility alone is not sufficient to induce significant clustering, even for very high area fractions of active particles (Fig.~\ref{fig:2}b). However, if the active particles experience a torque on colliding with passive ones, we observe the formation of clusters (Fig.~\ref{fig:2}c). Moreover, as in the experiments, the cluster phase corresponds to a nonequilibrium steady state (Fig.~S5)~\cite{SI}. The inclusion of torque between active and passive particles is motivated by the fact that bacteria are rod shaped, and therefore have a tendency to turn away from colloids on contact~\cite{lagarde2020colloidal} (Video S2)~\cite{SI}. Furthermore, we observe that the mean cluster size $\langle n\rangle=\sum nP(n)$ increases substantially with the area fraction $\phi_{\rm a}$ of active particles in the presence of torque ($\Gamma=25$/s), but remains nearly constant in the absence of torque (Fig.~\ref{fig:2}d).

\begin{figure}
  \centering
  \includegraphics[width=1\linewidth]{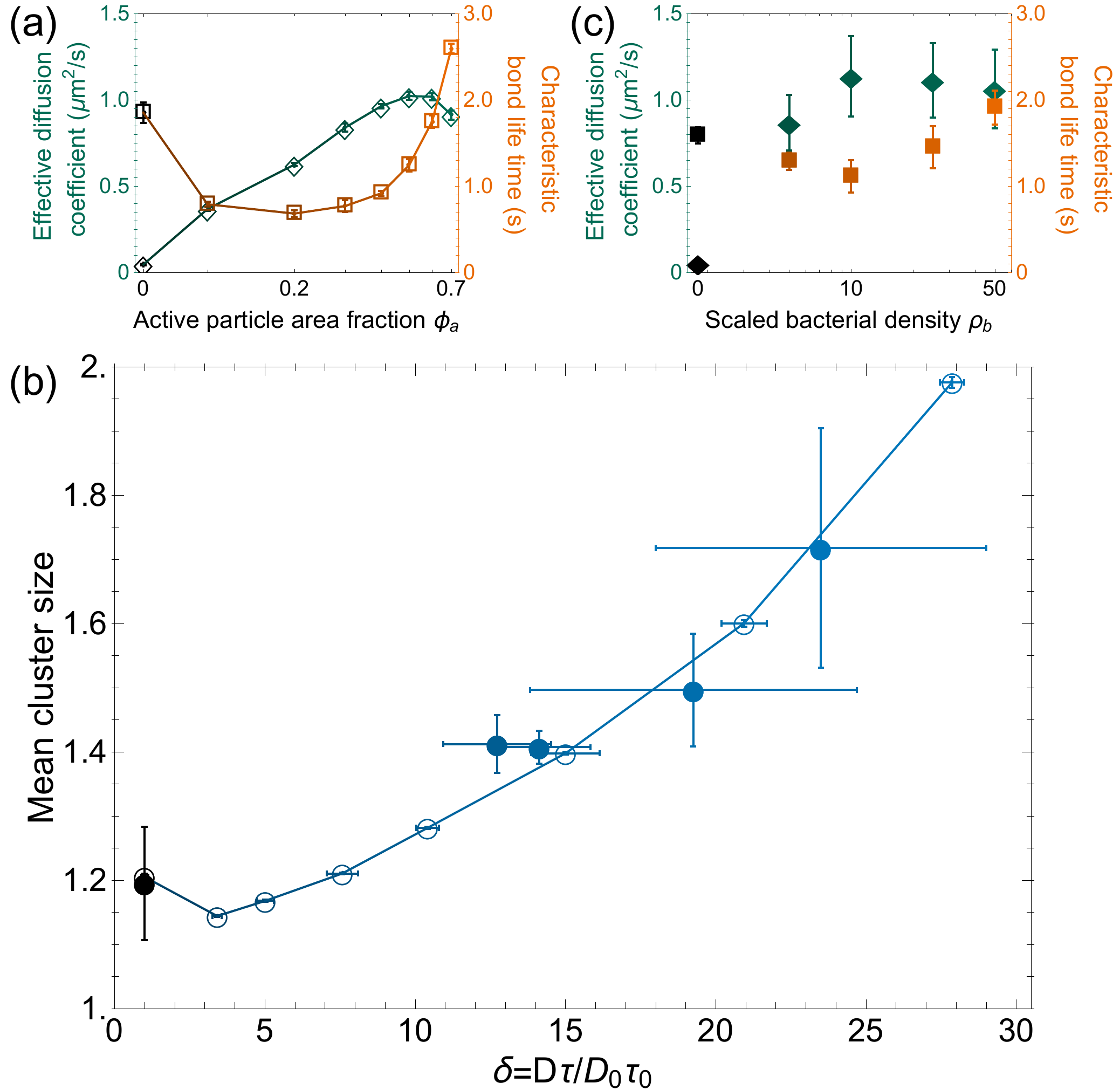}
  \caption{{\bf Increased effective attraction between passive particles leads to increased clustering in experiments and simulations.} (a) Single particle diffusion coefficient (open green diamonds) and characteristic bond lifetime (open orange squares) of passive particles as a function of the area fraction of active particles in simulations. (b) Mean cluster size as a function of the dimensionless parameter $\delta=D \tau/(D_0\tau_0)$ which quantifies the strength of effective attractions between passive particles, in simulations (open symbols) as well as experiments (filled symbols). (c) Single particle diffusion coefficient (filled green diamonds) and characteristic bond lifetime (filled orange squares) of colloids as a function of bacterial density in experiments.  In (a-c), black symbols correspond to results from controls without active particles {\it i.e.} $\phi_{\rm a}=0$ in simulations and $\rho_{\rm b}=0$ in experiments. Error bars in (a-c) are standard errors of the mean from three independent experiments or simulations and for some data points, error bars are smaller than the symbol size. }
  \label{fig:3}
\end{figure}

To gain further insights into the mechanism of clustering, we first investigated the dynamics of individual passive particles as well as the bonding kinetics of pairs of passive particles as a function of the area fraction of active particles $\phi_{\rm a}$ in simulations.  In the presence of aligning torques ($\Gamma>0$), active particles have two opposing effects on clustering. At the single particle level, collisions with active particles enhance diffusion of passive ones, which opposes clustering. However, the same collisions also result in effective attractions between pairs of passive particles, which ultimately leads to cluster formation. We quantified these competing effects using the single particle diffusivity $D$ (Fig.~\ref{fig:3}a, open green diamonds) and the mean pair bond lifetime $\tau$ (Fig.~\ref{fig:3}a, open orange squares) for passive particles (Fig.~S6-S7 and the section ‘Materials and Methods’ in supplemental materials)~\cite{SI}. Using the corresponding single particle diffusivity $D_0$ and mean pair bond lifetime $\tau_0$ under thermal fluctuations in the absence of active particles, we constructed a dimensionless parameter $\delta=D \tau/(D_0\tau_0)$ that serves as an indicator of the strength of effective attractions. In the absence of attractions, enhancement in diffusion would be accompanied by a corresponding decrease in bond lifetimes, resulting in $\delta \sim 1$. However, attractive interactions can lead to a substantial increase in lifetimes, leading to $\delta\gg 1$. Based on these physical considerations, we expect that larger values of $\delta$ should result in larger clusters, and this is indeed observed in our simulations (Fig.~\ref{fig:3}b, open symbols).

Finally, we examined whether the predicted increase in mean cluster size with $\delta$ can also be observed in experiments. The variation in diffusivity (Fig.~\ref{fig:3}c, filled green symbols) and mean pair bond lifetime (Fig.~\ref{fig:3}c, filled orange symbols) with bacterial density is qualitatively similar to that observed in simulations (Fig.~\ref{fig:3}a), particularly from moderate to high area fractions of active particles $\phi_{\rm a}$ (Fig.~S8-S9 and the section ‘Materials and Methods’ in supplemental materials)~\cite{SI}. Next we quantified the mean cluster size from our experimental data. While we hold the area fraction of passive colloids approximately constant ($\phi_{\rm p}=0.15$), in practice, the number of colloids in the field of view fluctuates across experiments, which influences our measurements of the mean cluster size. To account for these fluctuations, we define a corrected mean cluster size $\langle n_C^i\rangle=\frac{\langle N\rangle}{N_i}\langle n\rangle$, where $\langle n\rangle$ is the measured mean cluster size, $\langle N\rangle$ is the number of colloids in the field of view averaged across all experimental replicates and all bacterial densities, and $N_i$ is the number of colloids in the field of view for the i$^{\text{th}}$ experiment, averaged over the experimental duration. Both the bare as well as corrected mean cluster size increase with bacterial density (Fig.~S10)~\cite{SI}. Experiments with \textit{E. coli} also show qualitatively similar results (Fig.~S2)~\cite{SI}. The corrected mean cluster size plotted as a function of $\delta$ (Fig.~\ref{fig:3}b, filled symbols), is in agreement with the prediction from simulations, demonstrating that the clustering mechanism revealed by our simulations can adequately explain our experimental findings.

\begin{figure}
  \centering
  \includegraphics[width=1\linewidth]{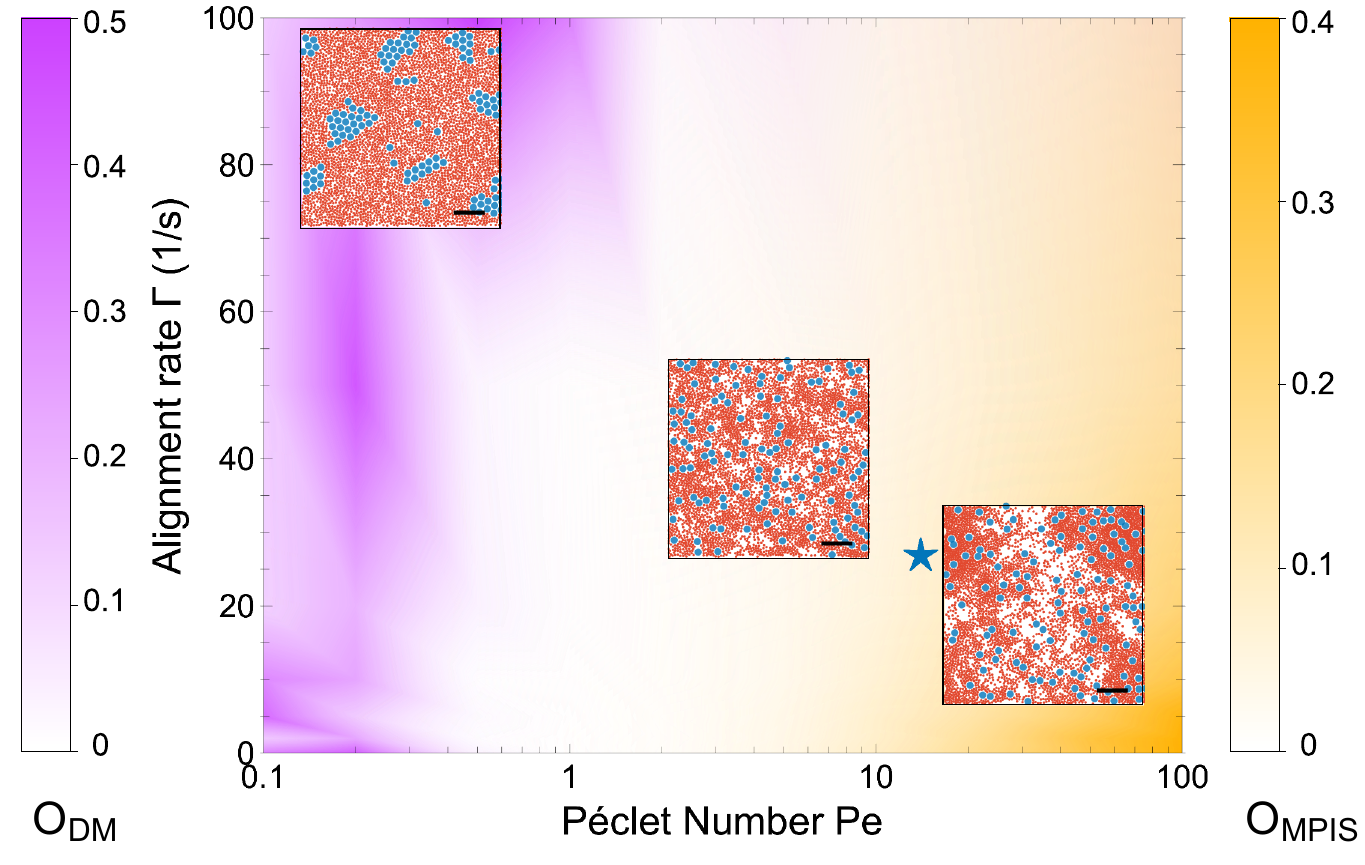}
  \caption{{\bf Numerical phase diagram of mixtures of active and passive particles in the P\'{e}clet number ($Pe$) - alignment rate ($\Gamma$) plane.}We change $Pe$ by varying $D_{\rm r}$ alone, and keeping other parameters constant. The system exhibits diffusivity-based phase separation at low $Pe$ (purple region). At intermediate $Pe$ (white region), the total density is homogeneous but passive particles exhibit steady state clustering. At high $Pe$, the system exhibits large density fluctuations consistent with MIPS (yellow region). The purple colormap (left) corresponds to the order parameter for the demixed phase ($O_{\rm DM}$) and yellow colormap (right) corresponds to the MIPS order parameter ($O_{\rm MIPS}$). Insets show representative simulation snapshots of the different phases, with passive particles in blue and active particles in red. The scale bar is $10 \mu$m. The blue star indicates parameters corresponding to our experimental data.} 
  \label{fig:4}
\end{figure}

To better understand the effect of aligning torques, and place our results in the context of prior work on mixtures of active and passive particles ~\cite{mccandlish2012spontaneous,stenhammar2015activity,dolai2018phase,grosberg2015nonequilibrium,weber2016binary}, we construct a numerical phase diagram for our system in the $Pe$-$\Gamma$ plane for fixed area fractions of passive ($\phi_{\rm p} = 0.15$) and active ($\phi_{\rm a} = 0.6$) particles. We delineate different phases using two order parameters. First we define an order parameter for the demixed regime $O_{\rm DM} = \langle N_L/N_p \rangle_{t}$, where $N_L$ is the number of particles in the largest cluster of passive particles, $N_p$ is the total number of passive particles, and $\langle \rangle_{t}$ indicates time averaging. We also define a MIPS order parameter $O_{\rm MIPS} = \langle \sigma_{\phi} \rangle_{t}/(\phi_{\rm a} + \phi_{\rm p})$, where $\sigma_{\phi}$ is the standard deviation of the local area fraction $\phi$, computed by dividing the simulation box into square cells of length $5d_{\rm a}$. As expected~\cite{grosberg2015nonequilibrium,weber2016binary}, we observe a demixed phase at low $Pe$ and motility-induced phase separation~\cite{cates2015motility} at high $Pe$ (Fig.~\ref{fig:4}, Fig.~11) ~\cite{SI}. Qualitatively, increasing $\Gamma$ shifts the demixed as well as MIPS phase boundaries to higher $Pe$ (Fig.~\ref{fig:4}), suggesting that aligning torques effectively reduce the persistence of active motion. 

Over a wide range of $Pe$ and $\Gamma$, we generically observe dynamic clustering. While there are no phase transitions within this regime, the mean cluster size increases with $\Gamma$ (Fig.~S12)~\cite{SI}. Using experimental measurements of bacterial speed ($v_b \approx 40 \mu$m/s), cell size ($d_b \approx 1.5 \mu$m), and tumbling time ($\approx 0.5$s), we estimate $Pe \approx 14$ and $\Gamma \sim v_b/d_b \approx 27$/s (Fig.~S4)~\cite{SI}. The motility parameters for our bacteria thus lie within the numerically predicted dynamic clustering regime (Fig.~\ref{fig:4}, blue star). Indeed, since most motile bacteria exhibit sizes $\sim 1\mu$m, speeds $\sim 10\mu$m/s and tumbling times $\sim 1$s, we expect dynamic clustering to be a generic feature of bacterial active matter.

In summary, we have shown that passive colloidal particles exhibit steady state dynamic clustering when immersed in a bath of motile bacteria and that the clustering can be tuned by changing the bacterial density (Fig.~\ref{fig:1}).
This effect can be understood as a balance between persistent active motion, which destabilizes clusters (Fig.~\ref{fig:2}b), and aligning torques, which stablizes them (Fig.~\ref{fig:2}c-d). Furthermore, dynamic clustering is a robust phenomenon observable in a broad parameter regime (Fig.~\ref{fig:4}). At the two body level, the stabilizing effect of aligning torques can be interpreted as an effective attraction between passive particles, as evidenced by the enhancement of bonding times relative to the expectation based on diffusion alone (Fig.~\ref{fig:3}b). The presence of such effective attractions suggests a possible route to direct and manipulate the self-assembly of building blocks by simultaneously tuning inter-particle interactions as well as spatiotemporal correlations in the active bath. Furthermore, using passive particles that break fore-aft~\cite{baek2018generic} or chiral~\cite{sokolov2010swimming} symmetry, it should be possible to self-assemble ordered structures with intrinsic translational or rotational dynamics, which cannot exist in equilibrium.

\acknowledgements S.G. was supported by a Human Frontier Science Program (HFSP) cross-disciplinary postdoctoral fellowship though Grant No. LT000470/2016-C. S.G. and A.S. acknowledge the Gordon and Betty Moore Foundation for support as Physics of Living Systems Fellows through Grant No. GBMF4513. This research was supported in part by the National Science Foundation under Grant No. NSF PHY-1748958 (to N.F.).

$^{\ddag}$S.G and J.L contributed equally to this work. 
\bibliography{references}

\end{document}

\title{Supplemental Information for \\Dynamic clustering of passive colloids in dense suspensions of motile bacteria}

\author{Shreyas Gokhale$^{\ddag,}$}
\affiliation{Department of Physics, Massachusetts Institute of Technology, Cambridge, MA 02139, USA}

\author{Junang Li$^{\ddag,}$}
\affiliation{Department of Physics, Massachusetts Institute of Technology, Cambridge, MA 02139, USA}

\author{Alexandre Solon}
\affiliation{Sorbonne Universit\'e, CNRS, Laboratoire de Physique Th\'eorique de la Mati\`ere Condens\'ee, LPTMC, F-75005 Paris, France}

\author{Jeff Gore}
\email[Corresponding author: ]{gore@mit.edu}
\affiliation{Department of Physics, Massachusetts Institute of Technology, Cambridge, MA 02139, USA}

\author{Nikta Fakhri}
\email[Corresponding author: ]{fakhri@mit.edu}
\affiliation{Department of Physics, Massachusetts Institute of Technology, Cambridge, MA 02139, USA}
 
\maketitle

\section*{Materials and Methods}

\subsection*{Bacterial strains, growth medium and conditions} 
\textit{Pseudomonas aurantiaca} bacteria (strain ATCC $\#$ 33663 with chromosomal insertion miniTn7-GmR-F1-06RFP for expression of red fluorescent protein (RFP)) were streaked from a frozen glycerol stock stored at -80 degree C onto plates containing 2$\%$ w/v Luria-Bertani (LB) medium and 1.5$\%$ w/v agar (both Beckton, Dickinson and Company, Franklin Lakes, NJ, USA). The plate was incubated for 24 hours at $30^{\circ} C$. To grow bacteria for microscopy experiments, individual colonies from the plate were used to inoculate 5 mL of growth medium containing 10 g/L Yeast Extract and 10 g/L Soytone (both Beckton, Dickinson and Company, Franklin Lakes, NJ, USA). Prior to inoculation, the growth medium was sterilized by passing through a 0.22 $\mu$m filter (VWR), and the pH was adjusted to 7 by dropwise addition of 100 mM sodium hydroxide (NaOH). Bacteria were grown for 16 hours using an orbital shaker at $30^{\circ} C$ and 250 rpm and allowed to reach saturation density ($\sim 3 \times 10^9$ colony forming units (CFU) per mL). The bacterial culture was concentrated by centrifuging at $4000g$ for three minutes. Different bacterial densities were achieved by removing different amounts of supernatant. The experiments with E. coli (Fig.~S\ref{figS2}) were performed using the strain M3K2R: W3110 containing the plasmid pSBIK3-RFP that confers resistance to kanamycin and enables expression of RFP. 

\subsection*{Sample and chamber preparation for microscopy experiments} Unfunctionalized silica colloids of diameter $\sigma=3.2\,\mu$m were purchased either from Bangs Laboratories, Inc. or microParticles GmbH. To demonstrate the interaction between individual Pseudomonas aurantiaca cells and colloids (Video~S2), we grew a fluorescent silica shell on the colloids that incorporated the fluorophore rhodamine B isothiocyanate (Sigma Aldrich) using the protocol devised by Kuijk et al.~\cite{1}. Video~S2 was captured using a customized fluorescence microscope with an EMCCD camera (ANDOR ixon 897) and 100X oil immersion objective lens (Nikon) at 20 frames per second (fps). Unfunctionalized silica colloids of diameter $5\,\mu$m, to be used as spacers, were purchased from Bangs Laboratories, Inc. The $\sigma=3.2\,\mu$m colloids were concentrated 2X by centrifugation and $5\,\mu$m silica colloids were diluted 10× in distilled deionized water. Prior to the experiments, both types of colloids were sonicated for 30 minutes, in order to break any clusters formed due to sedimentation during storage. 20 $\mu$L cconcentrated bacterial culture, 20 $\mu$L of 2X suspension of $3.2\,\mu$m colloids and 1 $\mu$L of 0.1X suspension of $5\,\mu$m colloids were mixed together by vortexing at 1500 rpm. 

In all experiments, 5 $\mu$L of this bacteria-colloid mixture was put between two polymer coverslips (ibidi GmbH, untreated), plasma cleaned at maximum radio frequency (RF) level for 15 minutes before use, filling an area of 18 mm × 18 mm. The polymer coverslips allow exchange of oxygen with the environment and help maintain bacterial motility over durations several times longer than those of our experiments.  The chamber was pressed by placing a known weight of 600g on top of the coverslips for $30$ s, which caused the $\sigma=5\,\mu$m colloids to stick to the top as well as bottom coverslip, thus serving the purpose of spacers. This arrangement ensures that the $\sigma=3.2\,\mu$m colloids experience quasi-2D confinement. Finally, the chamber was sealed with valap to prevent flows and evaporation. All experiments were performed in triplicate, using different bacterial colonies (biological replicates) and on separate days, to ensure that the observed phenomena are robust to fluctuations in bacterial density, motility and metabolism. 

\subsection*{Optical microscopy set-up} 
The samples were imaged in bright field with a 40X or 60X air objective on a Motic AE2000 inverted microscope. Videos were acquired for 5 minutes at 6 fps using a CMOS digital camera (AmScope MU500). Prior to capturing videos, we waited for 15 minutes to allow any incidental bulk flows in the sample chamber to subside. The field of view typically contained $\sim 600 \pm 50$ colloids.

\subsection*{Testing for colloidal clustering due to growth medium conditioned by bacteria}
\begin{figure*}
\renewcommand{\figurename}{Fig. S\hspace{-0.3em}}
  \centering
  \includegraphics[width=1\linewidth]{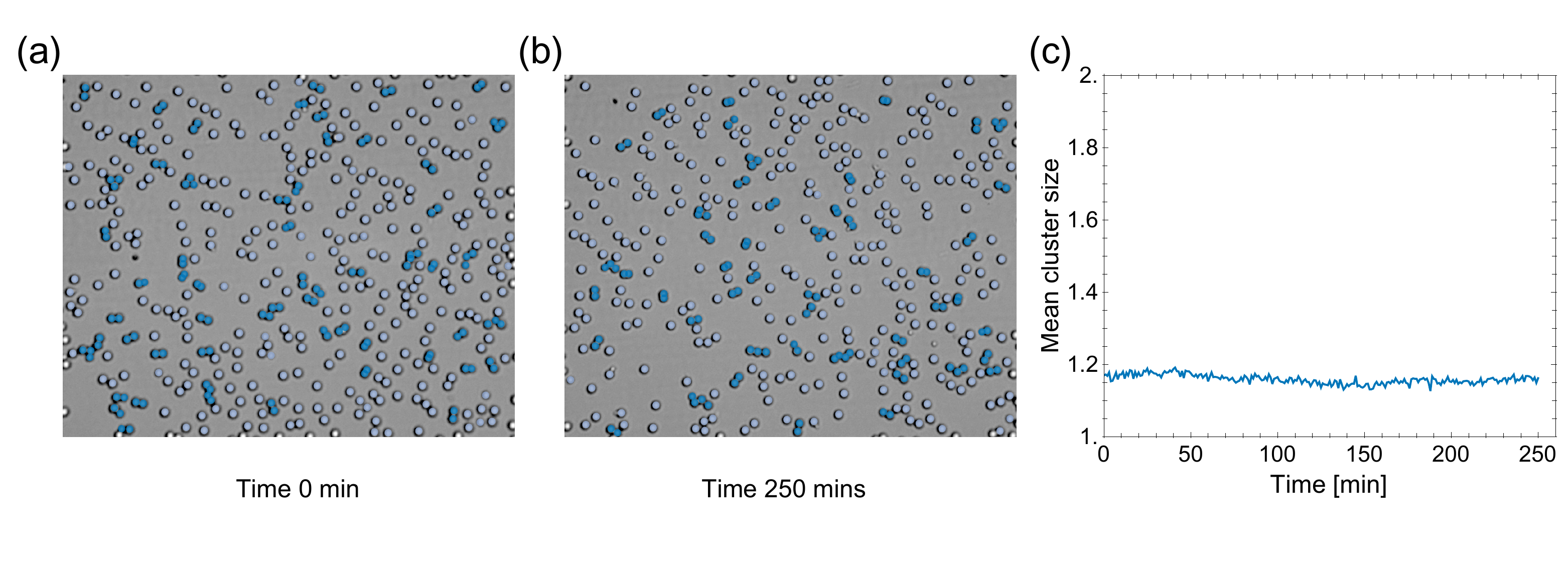}
  \caption{{\bf Control experiment with passive colloids immersed in bacterial supernatant.} A snapshot of the colloid-supernatant mixture at the beginning of our control experiment is shown in (a) and a shapshot taken $250$ minutes later is shown in (b). Clusters with two or more bonded colloids are shown in dark blue while monomers are shown in light blue. (c) Plot of the mean cluster size as a function of time. The passive colloids do not exhibit any tendency towards clustering even after $250$ minutes, which is 50 times longer than the typical duration of our experiments ($\sim 5$ minutes).}
  \label{figS1}
\end{figure*}

\begin{figure*}
\renewcommand{\figurename}{Fig. S\hspace{-0.3em}}
  \centering
  \includegraphics[width=1\linewidth]{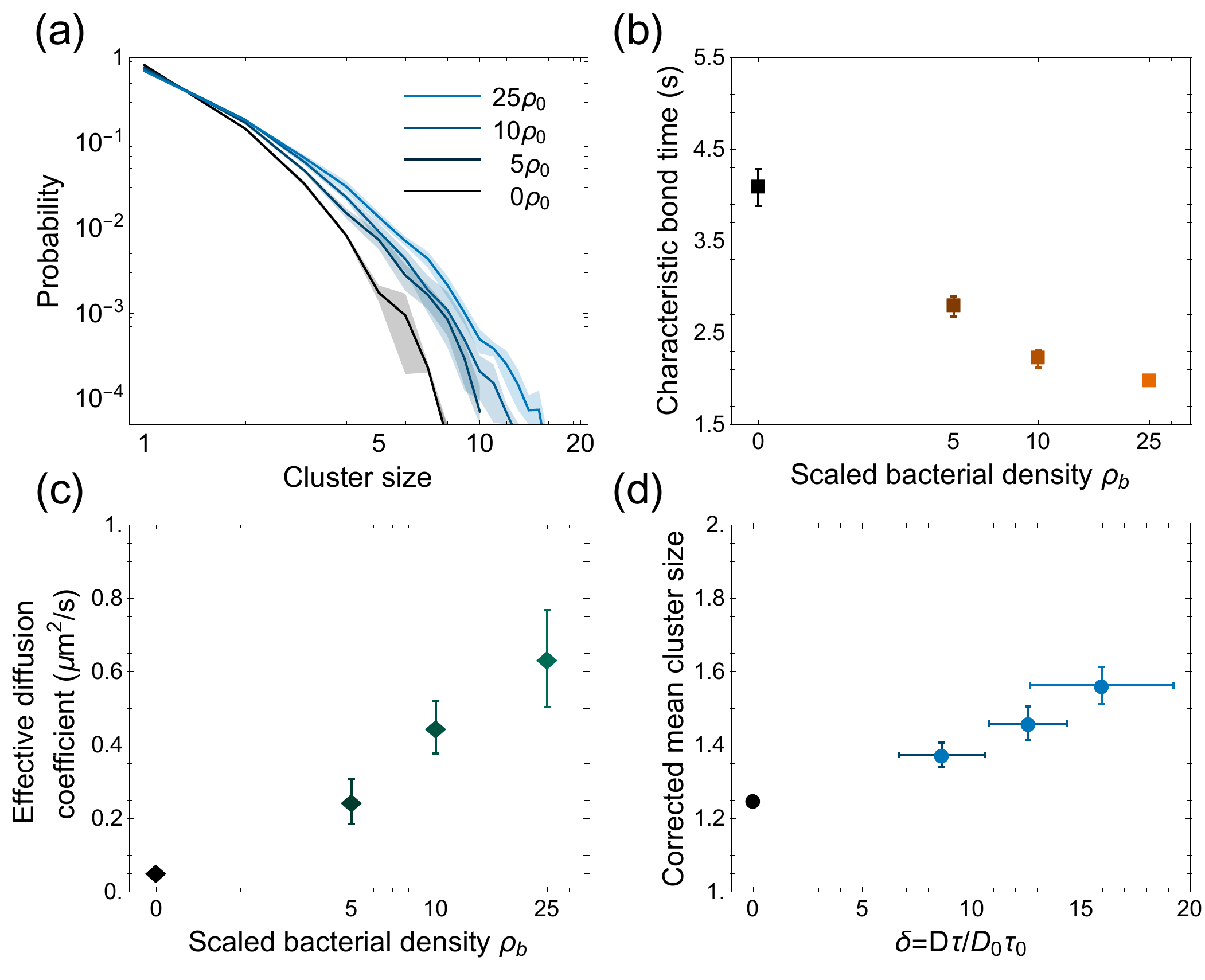}
  \caption{{\bf Results of experiments with \textit{E.coli}.} (a) Cluster size distribution $P(n)$ for various bacterial densities. The saturation density $\rho_0$ for \textit{E. coli} is $\sim 7 \times 10^9$ CFU / mL. (b-c) Characteristic bond lifetime (b) and effective diffusion coefficient (c) of colloids as a function of bacterial density. (d) Corrected mean cluster size as a function of the dimensionless parameter $\delta$. Shaded region in (a), and error bars in (b-d) are standard errors of the mean across three independent experiments.}
  \label{figS2}
\end{figure*}

To ensure that the growth medium conditioned by bacteria does not lead to aggregation of colloids due to salts or other dissolved molecules, we performed a control experiment. After concentrating the bacteria, we filtered the supernatant containing the conditioned growth medium through a 0.2 $\mu$m syringe filter (Pall Corporation) to remove bacterial cells. We mixed 20 $\mu$L of this filtered supernatant with 20 $\mu$L of  2X suspension of 3 mm colloids and 1 $\mu$L of 0.1X suspension of $\sigma=5\,\mu$m colloids and loaded 5 $\mu$L of this sample into the same type of imaging chamber as described above. These conditions are identical to those in our main experiments, with the exception that the bacterial suspension is replaced by the conditioned growth medium. We then took a time lapse video at 1 frame per minute for 250 minutes, which is 50 times longer than the duration of our experiments. We observed no evidence of colloidal clustering over this period (Fig.~S\ref{figS1}), showing that the clustering observed in our experiments is due to bacterial motility alone, and not due to chemical modifications in the growth medium that occur during bacterial growth.

\subsection*{Colloid tracking and cluster identification} Colloids were localized and tracked using Blair and Dufresne’s publicly available Matlab version of the particle tracking code originally developed by Crocker and Grier~\cite{2}.  Particle trajectories were de-drifted to remove the effects of residual bulk flows in the sample as well as any drift in the microscope stage. Clusters were identified using a distance cutoff. Two colloids separated by less than 1.2$\sigma$, where $\sigma$ is the colloid diameter, were classified as belonging to the same cluster.

\subsection*{Measurement of bond lifetimes}
We assume that two colloids are bonded to each other if their centroids are separated by a distance smaller than 1.2$\sigma$, the same cutoff distance used for clustering analysis. A bond lifetime for a pair of colloids is defined as the duration over which two colloids remain separated by a distance smaller than 1.2$\sigma$. The bond is said to be broken once the separation exceeds this cutoff distance. Two colloids can thus form and break bonds with each other as well as with other colloids, several times during the experiment. While bond lifetimes can help us quantify effective pairwise interactions, these can also be affected by clustering, since particles in the interior of clusters will typically have longer bond lifetimes. Thus, clustering leads to heavy tails in the bond lifetime distributions (Fig.~S\ref{figS8}). We ignore these heavy tails and only consider the initial exponential decay of bond lifetimes in order to get a more accurate estimate of the characteristic bond lifetime of pairs of colloids. The characteristic bond lifetime is defined as the decay constant of the exponential fit to the initial decay of the bond lifetime distribution. In simulations, we compute the bond lifetime by simulating two passive particles diffusing in a bath of active particles, to avoid possible many body effects. 

\subsection*{Measurement of effective diffusion coefficient} To extract the effective diffusion coefficient of colloids in our experiments, we calculated the mean square displacement (MSD) using a time average for a single track as well as an ensemble average over the tracks of all colloids in the field of view. While the diffusion coefficient is a property of individual colloids, its measurement can be affected by the clustering observed in our experiments. To remove the effect of clustering on diffusivity measurements, we only considered those particle trajectories that had on average less than one nearest neighbor over their duration. Nearest neighbors were defined using the same distance cutoff used to identify clusters. The MSDs for all bacterial densities studied were diffusive at long times (Fig.~S\ref{figS9}), which allowed us to extract the diffusion coefficient of colloids. In simulations, we extract the diffusion coefficient by simulating a single passive particle in a bath of active particles and averaging over time.

\section*{Supplemental Video Captions}

\noindent \textbf{Supplemental Video 1:} Video showing dynamic clustering of colloids inside a bacterial bath of density $\rho_{\rm b} = 50\rho_0$. The duration of the video is $\approx 1$ minute with a frame rate of $6$ fps and the scale bar is $10\, \mu$m.
\\
\\
\textbf{Supplemental Video 2:} Video showing a collision between a colloid and a \textit{P. aurantiaca} cell. Fluorescence microscopy video of a single bacterium colliding with a colloid. Yellow dashed circle outlines the bacterial cell. The pronounced change in direction of the bacterium’s motion provides a clear indication of the existence the orientational interactions. The scale bar is $5\,\mu$m.

\newpage

\begin{figure*}
\renewcommand{\figurename}{Fig. S\hspace{-0.3em}}
  \centering
  \includegraphics[width=1\linewidth]{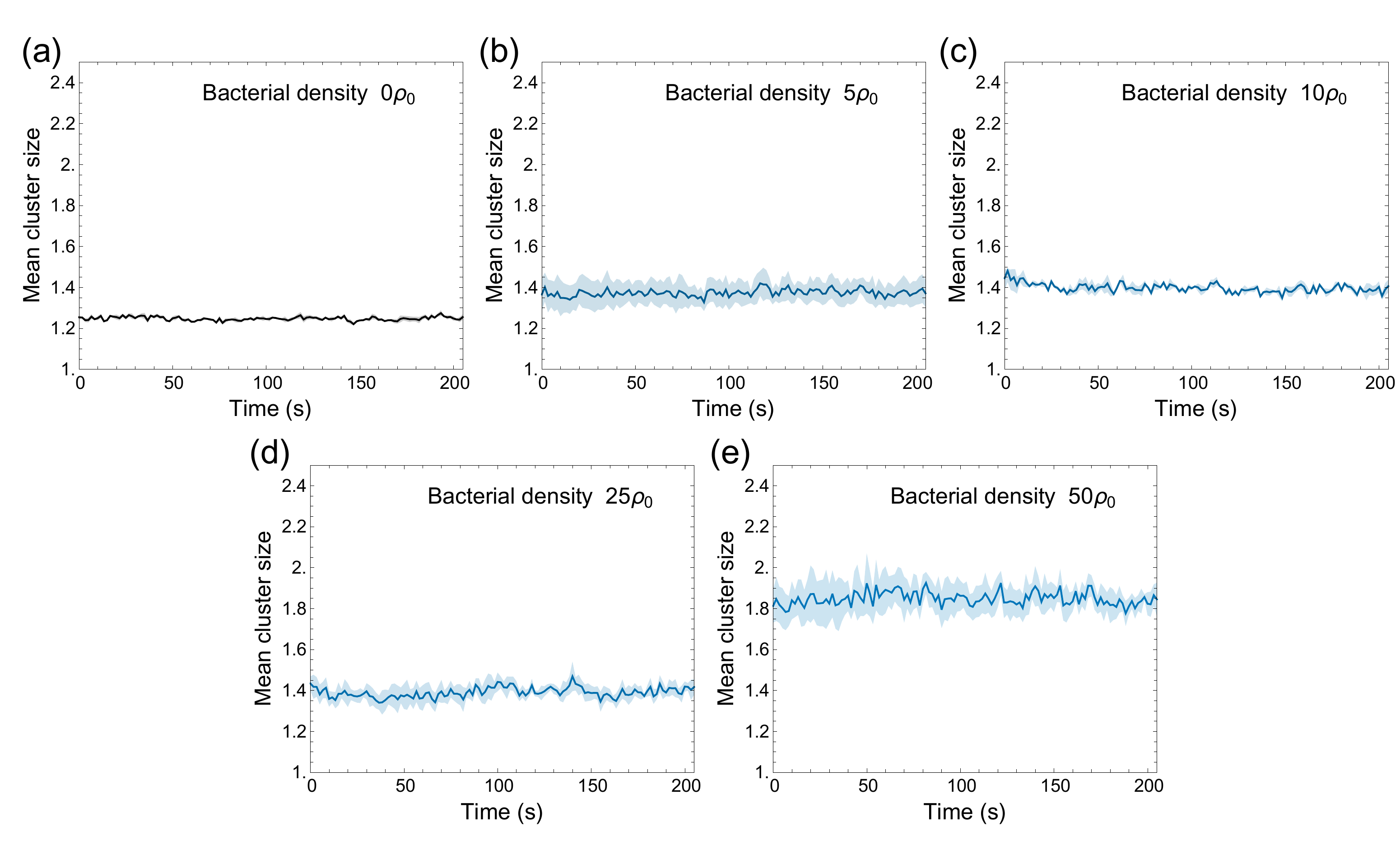}
  \caption{{\bf Time evolution of the mean cluster size in experiments.} Mean cluster size as a function of time for bacterial densities $0\rho_0$ (control) (a), $5\rho_0$ (b), $10\rho_0$ (c), $25\rho_0$ (d) and $50\rho_0$ (e). Shaded regions are standard errors of the mean across three independent experiments. In all the experiments, mean cluster size fluctuates about a constant value but does not drift or change abruptly over time, which demonstrates that our system is in steady state.}
  \label{figS3}
\end{figure*}

\begin{figure*}
\renewcommand{\figurename}{Fig. S\hspace{-0.3em}}
  \centering
  \includegraphics[width=1\linewidth]{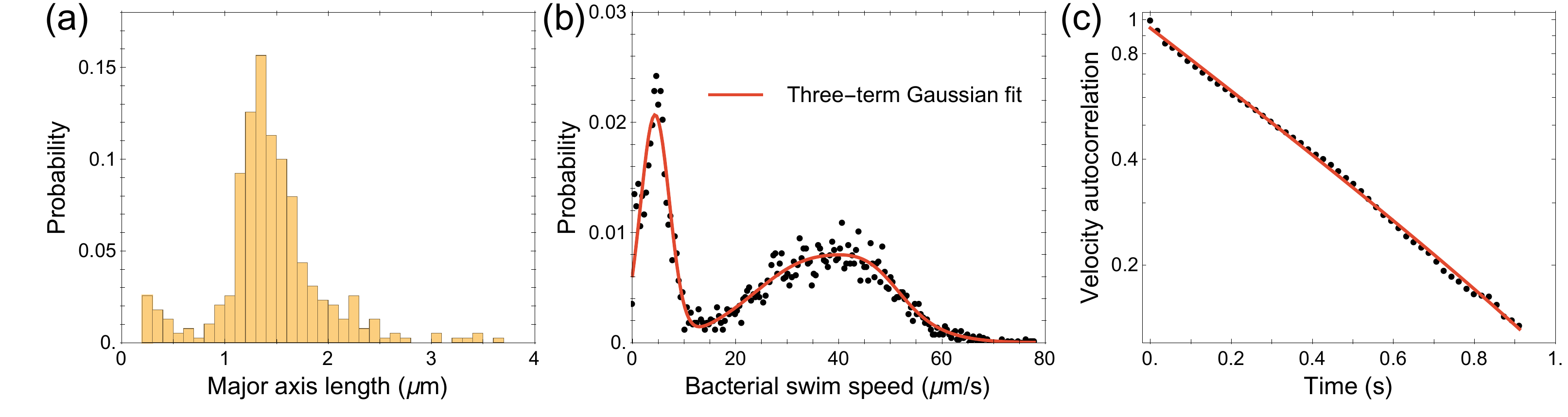}
  \caption{{\bf Measurement of P\'{e}clet number for \textit{P. aurantiaca} bacteria.} The bacterial size, speed and tumbling time were computed using optical microscopy videos captured at $60$X magnification and frame rate of 53.77 fps. (a) Distribution of bacterial body length, corresponding to the major axis of the best fit ellipse to the bacterial intensity profile. From the mean and standard deviation of the distribution, we get bacterial size $d_b = 1.5 \mu$m $\pm 0.3\mu$m. (b) Experimentally measured bacterial swim speed distribution (black dots). Due to occasional sticking or non-viability of bacterial cells, the distribution has two peaks corresponding to motile and non-motile cells. To obtain the location and width of the second peak corresponding to motile cells, we fit a model with three Gaussian terms in MATLAB (red line). From the peak location and full width at half maximum, we obtain the bacterial speed $v_b = 41 \mu$m$/$s $\pm 16 \mu$m$/$s. (c) To compute the tumbling time, we compute the normalized bacterial velocity autocorrelation $\langle \vec{v_b(t}) \cdot \vec{v_b(0)} \rangle/\langle \lvert \vec{v_b} \rvert^2 \rangle$. From an exponential decay fit (red line) to the experimental data, we obtained the tumbling timescale to be $0.5$ s $\pm 0.02$ s. From the data in (a-c), we computed the P\'{e}clet number for our system to be $Pe = 13.6 \pm 5.7$.}
  \label{figS4}
\end{figure*}

\begin{figure*}
\renewcommand{\figurename}{Fig. S\hspace{-0.3em}}
  \centering
  \includegraphics[width=1\linewidth]{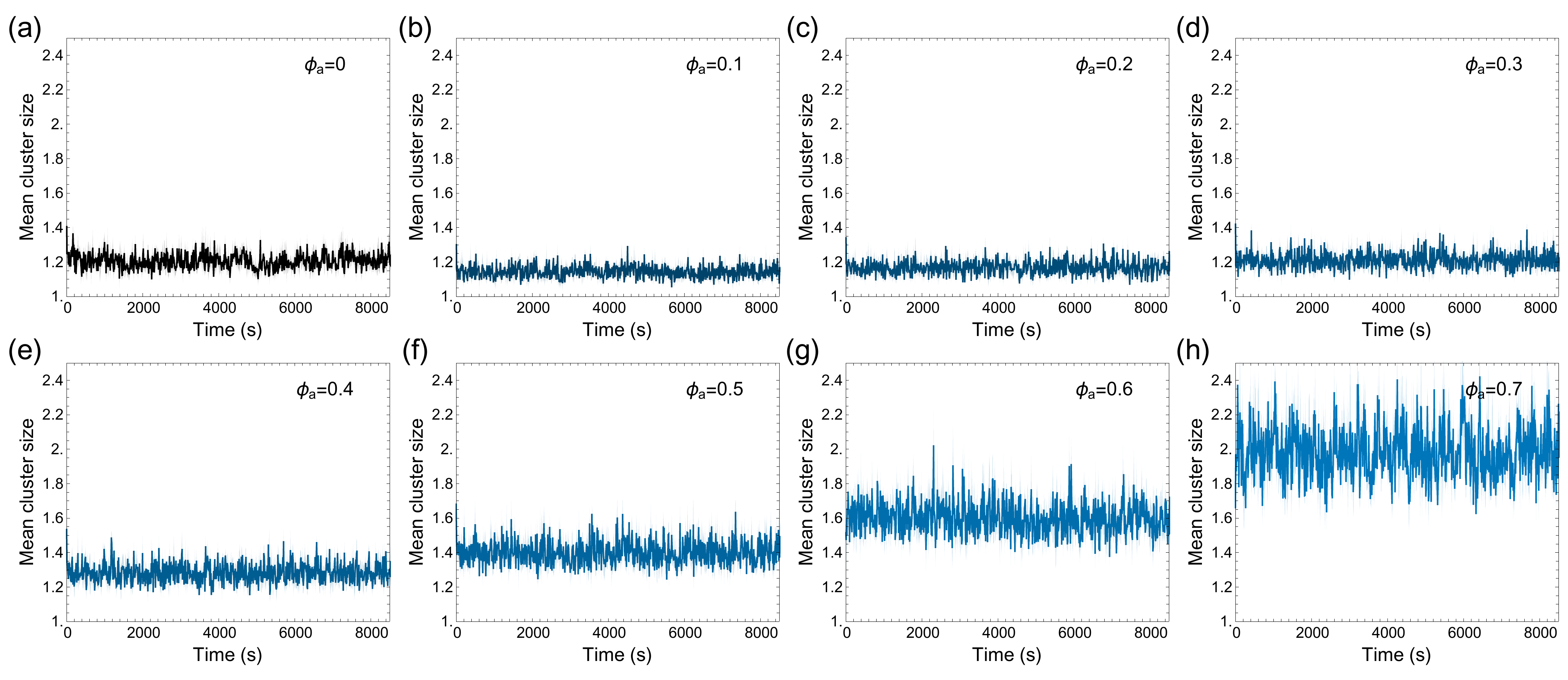}
  \caption{{\bf Time evolution of the mean cluster size in simulations.} Mean cluster size as a function of time for active particle area fraction $\phi_{\rm a} = 0$ (a), $0.1$ (b), $0.2$ (c) $0.3$ (d), $0.4$ (e), $0.5$ (f) $0.6$ (g) and $0.7$ (h). The mean cluster size is constant in time, indicating that the system is in a nonequlibrium steady state. The area fraction of passive particles is $\phi_{\rm p} = 0.15$}
  \label{figS5}
\end{figure*}

\begin{figure*}
\renewcommand{\figurename}{Fig. S\hspace{-0.3em}}
  \centering
  \includegraphics[width=1\linewidth]{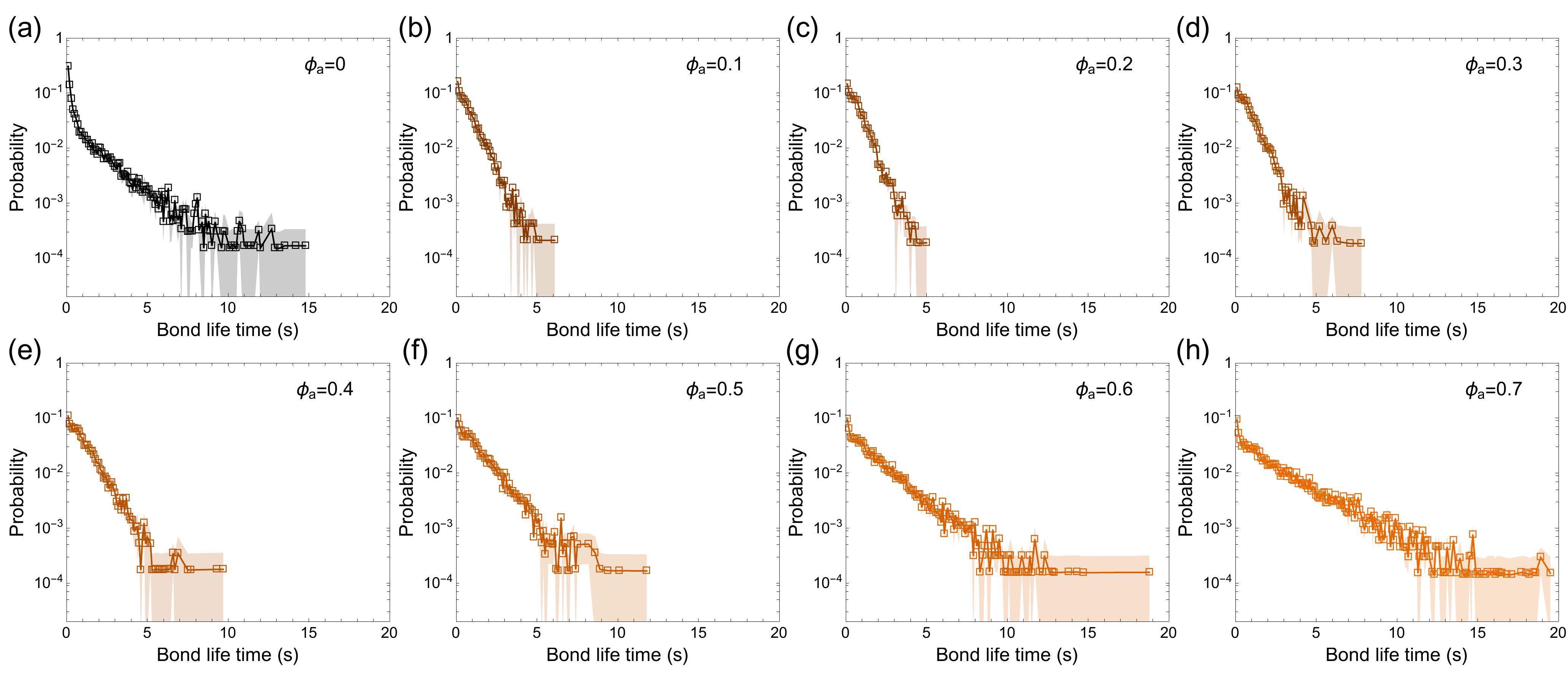}
  \caption{{\bf Bond lifetime distribution for pairs of passive particles in simulations.} Bond lifetime distributions corresponding to active particle area fraction $\phi_{\rm a} = 0$ (a), $0.1$ (b), $0.2$ (c) $0.3$ (d), $0.4$ (e), $0.5$ (f) $0.6$ (g) and $0.7$ (h). Shaded regions are the standard errors of the mean across three independent simulations.}
  \label{figS6}
\end{figure*}

\begin{figure*}
\renewcommand{\figurename}{Fig. S\hspace{-0.3em}}
  \centering
  \includegraphics[width=1\linewidth]{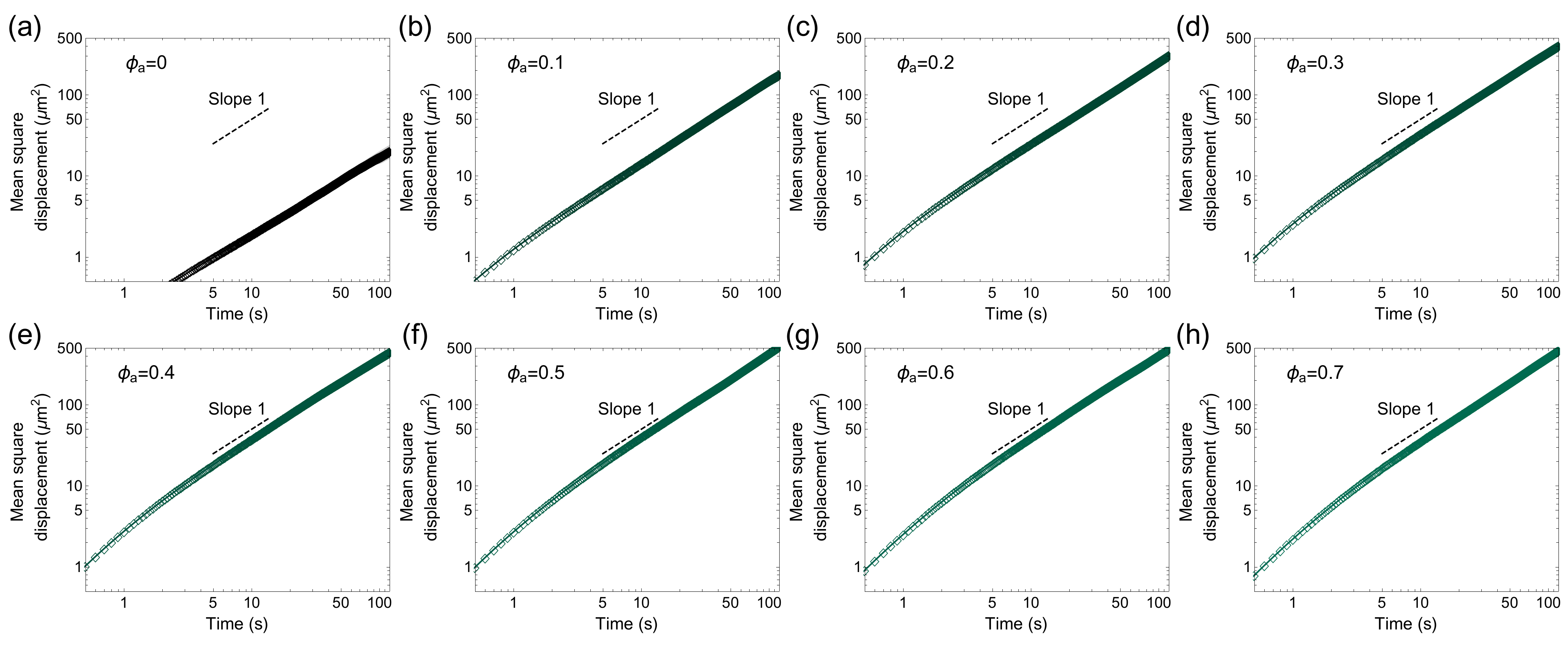}
  \caption{{\bf Mean squared displacement of passive particles in simulations.} Mean squared displacements of passive particles for active particle area fraction $\phi_{\rm a} = 0$ (a), $0.1$ (b), $0.2$ (c) $0.3$ (d), $0.4$ (e), $0.5$ (f) $0.6$ (g) and $0.7$ (h). Shaded regions are the standard errors of the mean across three independent simulations.}
  \label{figS7}
\end{figure*}

\begin{figure*}
\renewcommand{\figurename}{Fig. S\hspace{-0.3em}}
  \centering
  \includegraphics[width=1\linewidth]{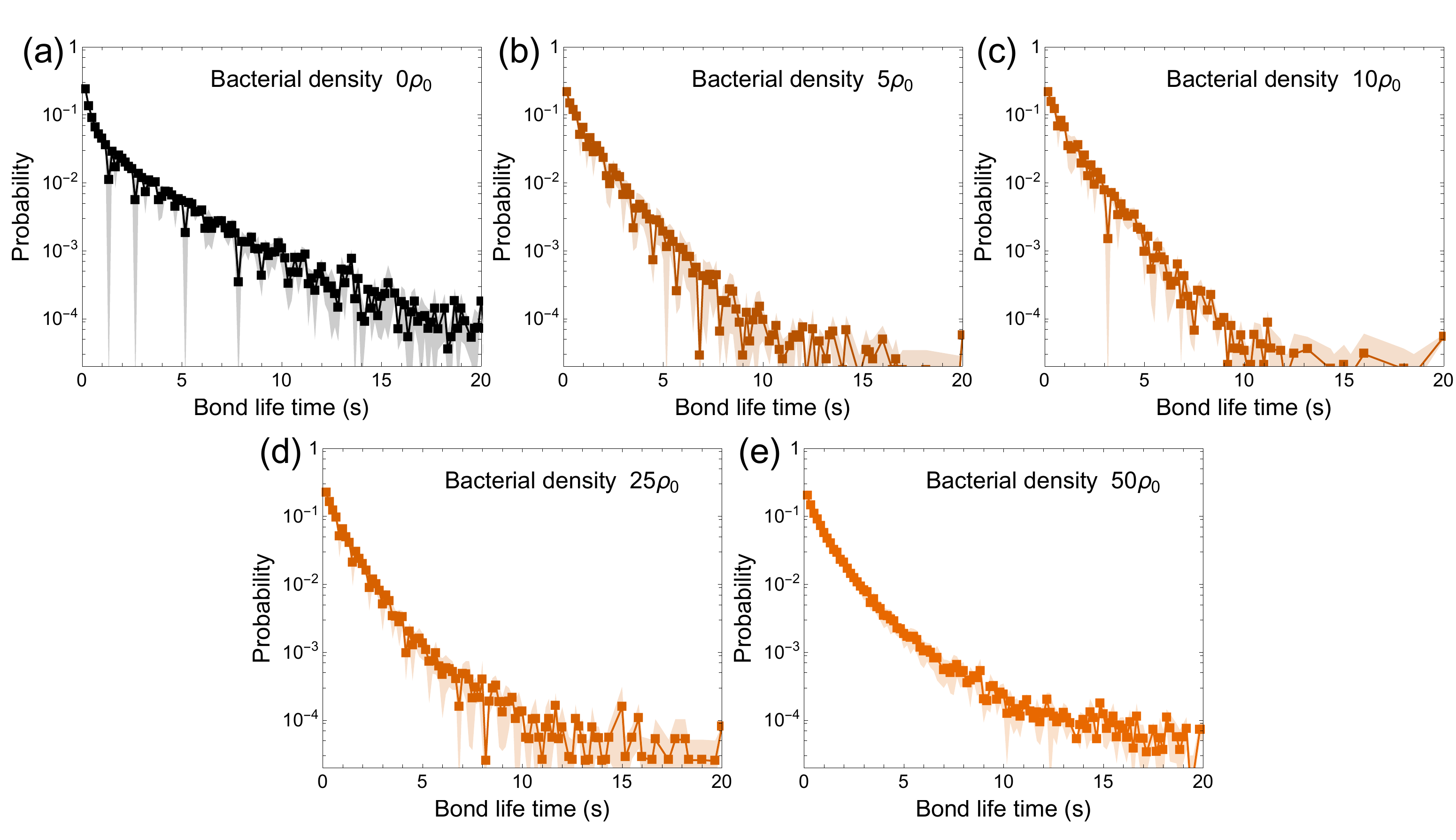}
  \caption{{\bf Experimental bond lifetime distributions for pairs of passive colloids.} Bond lifetime distributions corresponding to bacterial densities $0\rho_0$ (control) (a), $5\rho_0$ (b), $10\rho_0$ (c), $25\rho_0$ (d) and $50\rho_0$ (e). Shaded regions are the standard errors of the mean across three independent experiments.}
  \label{figS8}
\end{figure*}

\begin{figure*}
\renewcommand{\figurename}{Fig. S\hspace{-0.3em}}
  \centering
  \includegraphics[width=1\linewidth]{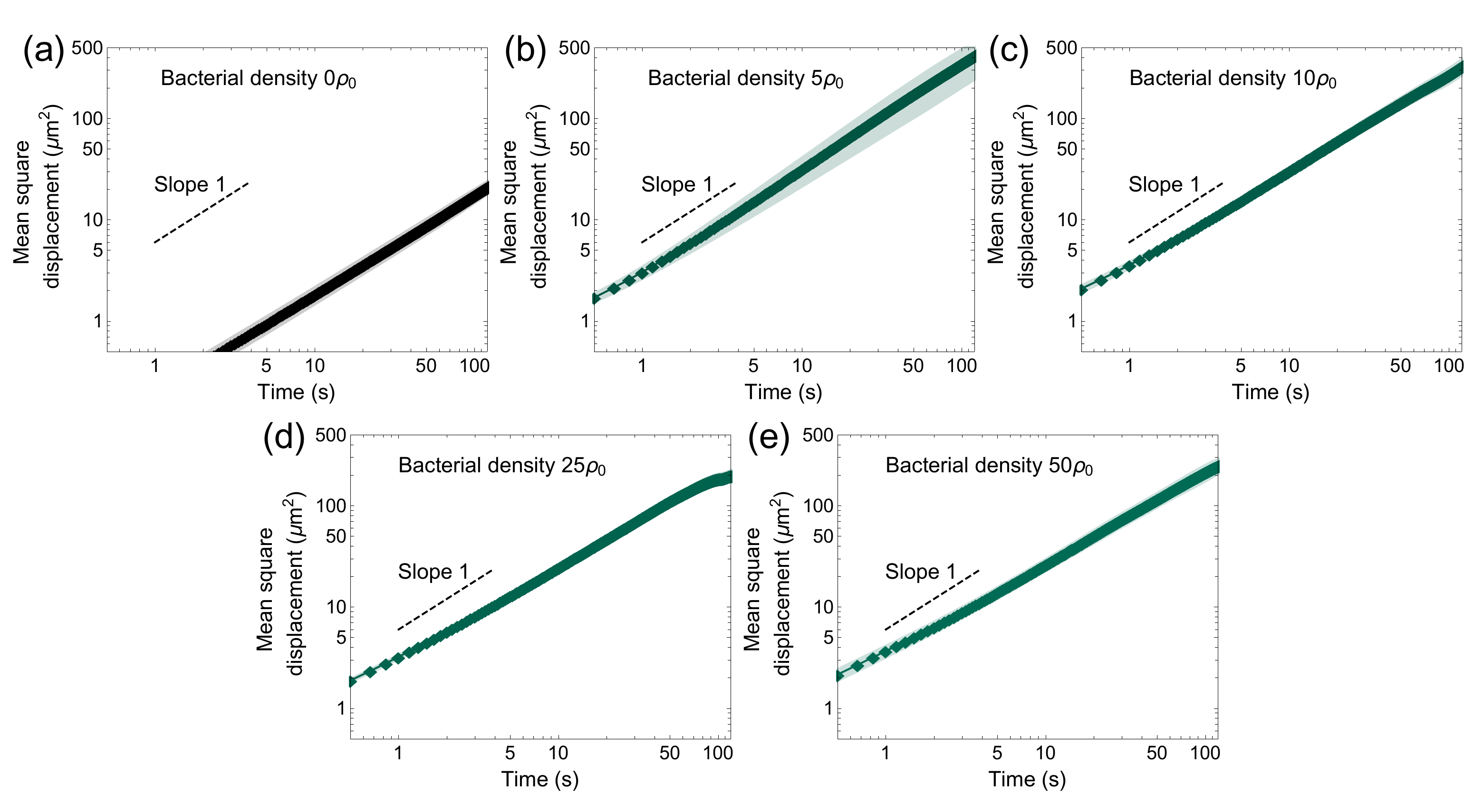}
  \caption{{\bf Mean squared displacement of colloids.} Mean squared displacement of colloids for bacterial densities $0\rho_0$ (control) (a), $5\rho_0$ (b), $10\rho_0$ (c), $25\rho_0$ (d) and $50\rho_0$ (e), showing diffusive dynamics at long times. Shaded regions are the standard errors of the mean across three independent experiments.}
  \label{figS9}
\end{figure*}

\begin{figure*}
\renewcommand{\figurename}{Fig. S\hspace{-0.3em}}
  \centering
  \includegraphics[width=1\linewidth]{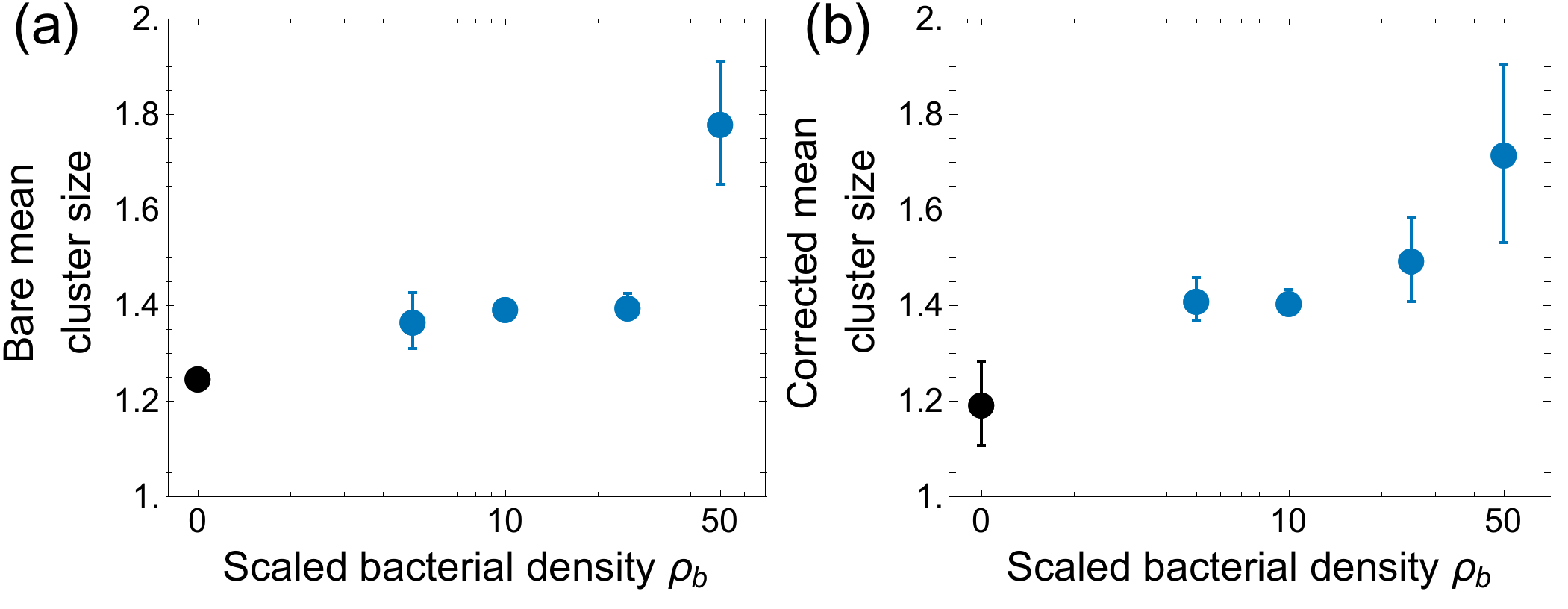}
  \caption{{\bf Bare and corrected mean cluster size in experiments.} Bare (a) and corrected (b) mean cluster size as a function of scaled bacterial density $\rho_b$. Error bars are standard errors of the mean across three independent simulations.}
  \label{figS10}
\end{figure*}

\begin{figure*}
\renewcommand{\figurename}{Fig. S\hspace{-0.3em}}
  \centering
  \includegraphics[width=1\linewidth]{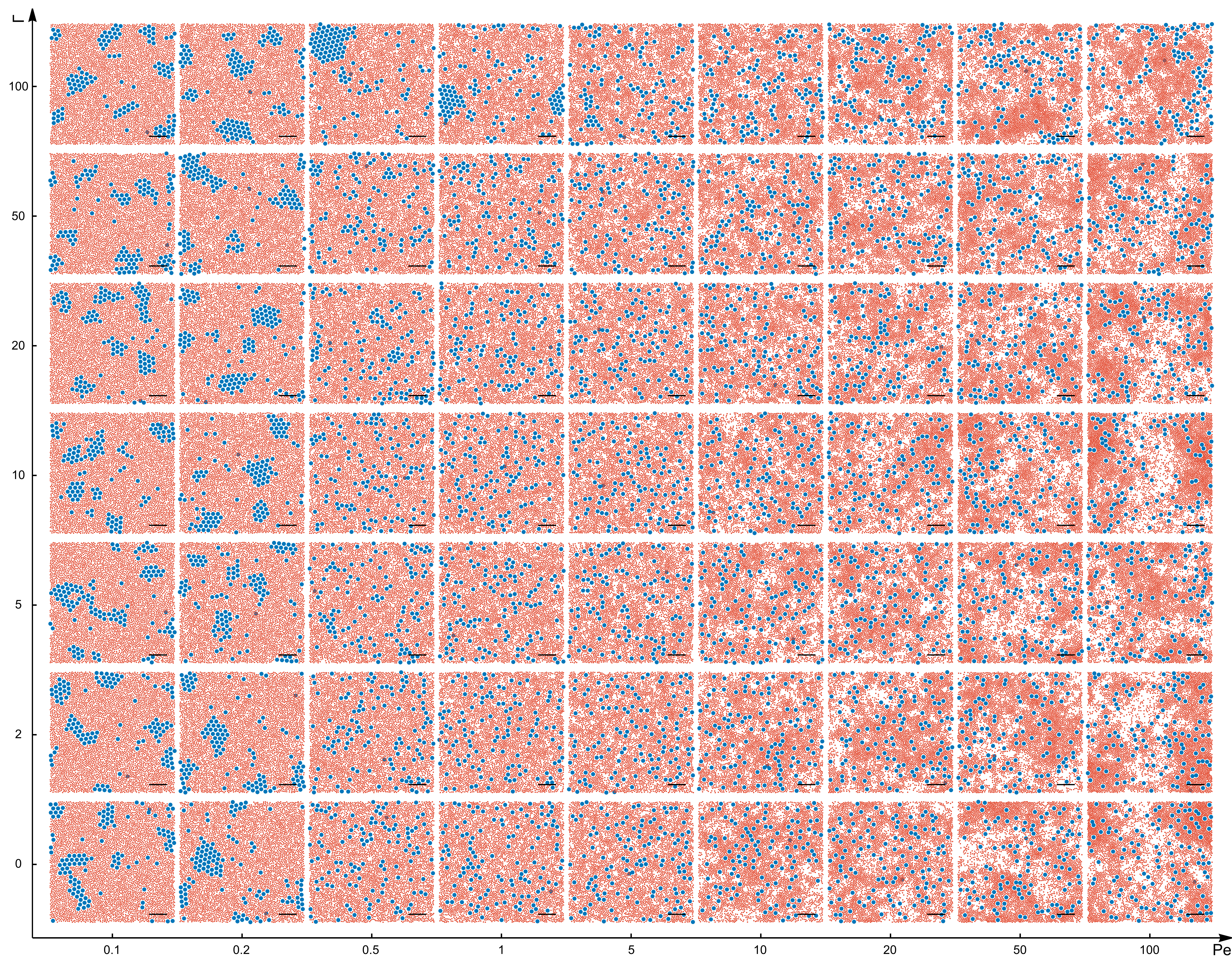}
  \caption{{\bf Visual phase diagram in the $Pe - \Gamma$ plane.} Snapshots from simulations used to construct the phase diagram in Fig. 4 of the main text. Active particles are shown in red and passive ones in blue. Demixed phase at low $Pe$ is characterized by phase separation of active and passive particles, while MIPS phase at high $Pe$ is characterized by large density fluctuations. In all simulations, area fraction of active particles $\phi_{\rm a} = 0.6$ and area fraction of passive particles $\phi_{\rm p} = 0.15$. The scale bar is $10 \mu$m.}
  \label{figS11}
\end{figure*}

\begin{figure*}
\renewcommand{\figurename}{Fig. S\hspace{-0.3em}}
  \centering
  \includegraphics[width=1\linewidth]{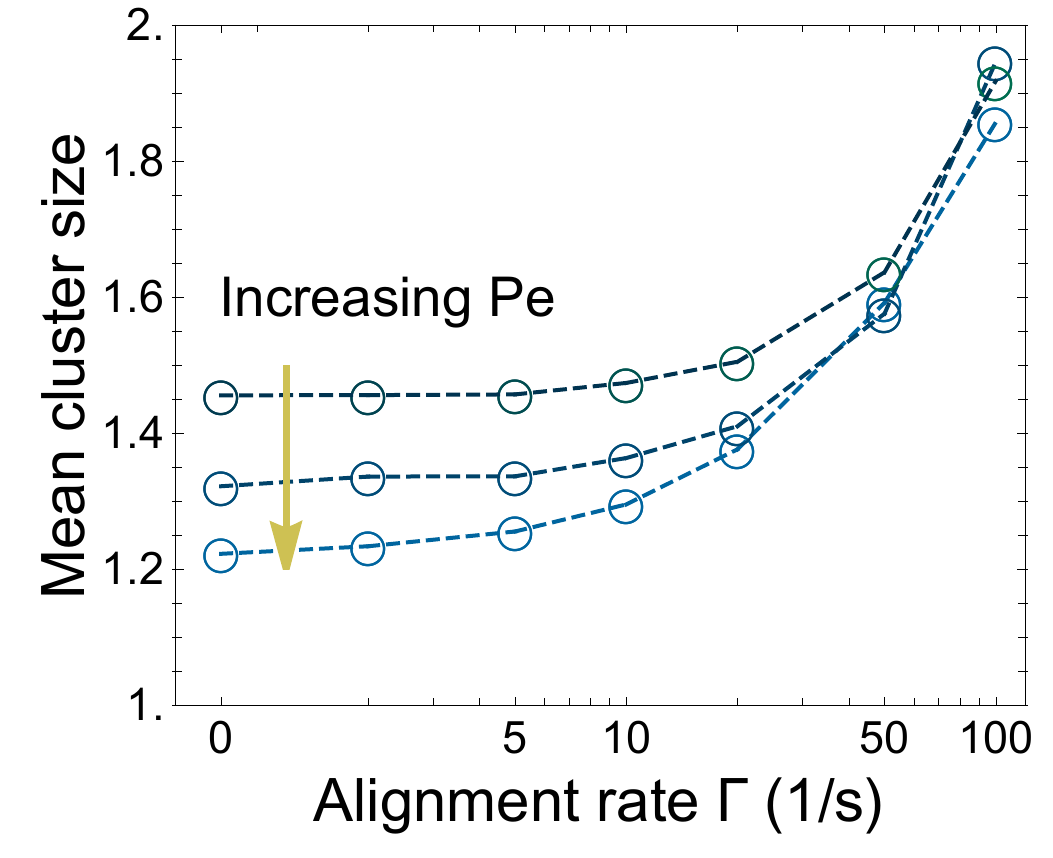}
  \caption{{\bf Dependence of cluster size on $\Gamma$.} Mean cluster size as a function of the alignment rate $\Gamma$ for $Pe = 1$ (top), $Pe = 2$ (middle) and $Pe = 10$ (bottom). The area fraction of active particles $\phi_{\rm a} = 0.6$ for all plots.}
  \label{figS12}
\end{figure*}